# Variability and Fidelity Limits of Silicon Quantum Gates Due to Random Interface Charge Traps

Tong Wu and Jing Guo

*Abstract*—Silicon offers an attractive material platform for hardware realization of quantum computing. In this study, a microscopic stochastic simulation method is developed to model the effect of random interface charge traps in silicon metal-oxide-semiconductor (MOS) quantum gates. The statistical results show that by using a fast two-qubit gate in isotopically purified silicon, the two-qubit silicon-based quantum gates have the fidelity >98% with a probability of 75% for the state-of-the-art MOS interface quality. By using a composite gate pulse, the fidelity can be further improved to >99.5% with the 75% probability. The variations between the quantum gate devices, however, are largely due to the small number of traps per device. The results highlight the importance of variability consideration due to random charge traps and potential to improve fidelity in silicon-based quantum computing.

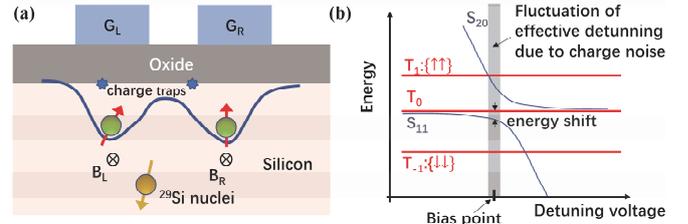

Fig. 1. (a) Schematic sketch of the modeled silicon MOS two-qubit entangling quantum gates device. The left and right gates ($G_L$ and $G_R$ respectively) define two spin quantum dots. $B_L$ and $B_R$ are the external magnetic field applied on the left and right quantum dots, respectively. The random interface charge traps result in charge noise decoherence, and $^{29}$Si nuclei spins result in decoherence. (b) Operation of the controlled phase gates by applying a detuning gate voltage $\Delta V_g$. The singlet S$_{11}$ and triplet (T$_1$, T$_0$, and T$_{-1}$) energy levels as a function of $\Delta V_g$ are shown. A main effect of the random charge trap noise is perturbation of the effective detuning voltage around the bias point, as schematically shown by the vertical gray bar.

## I. INTRODUCTION

Among various hardware realization of quantum computing, silicon-based quantum computing is particularly attractive for its potential to achieve high integration density, low cost, and compatibility with the mainstream IC technologies [1][2][3][4]. One challenge of silicon-based quantum computing is how to realize entangling quantum gates with high fidelity [5][6][7]. To achieve strong interaction and entanglement between neighboring qubits, it is necessary to have scaled devices[8], of which one important issue is device-to-device variability [9]. To integrate physical qubits to a quantum computing system, it is necessary to understand and control device-to-device variability for silicon-based quantum gates.

In this letter, the variability and fidelity limits of two-qubit entangling gates based on a metal-oxide-semiconductor (MOS) structure are investigated [5][10][11][12]. A microscopic stochastic device simulation method is developed to treat the effect of random charge traps in MOS interface and nuclei spin dephasing. The results indicate that for a well-designed device, which takes advantage of long spin coherence time in isotopically purified silicon and fast gate time, high gate fidelity values can be achieved for the state-of-the-art MOS interface quality. The variability of the fidelity between devices, however, is large due to the randomness of the interface charge traps and the small size of the device. The results indicate the importance of variability due to random charge traps and potential to achieve high fidelity in the silicon MOS platform for realizing quantum computing technology.

This work was supported in part by NSF Award 2007200.
The authors are with the Department of Electrical and Computer Engineering, University of Florida, Gainesville, FL 32611-6130 USA (e-mail: guoj@ufl.edu).

## II. APPROACH

The modeled silicon MOS two-qubit entangling quantum gates device is schematically shown in Fig. 1(a). The two spin qubits are defined by the applied gate voltages on two gates, $G_L$ and $G_R$. The externally applied magnetic fields, $B_L$ and $B_R$, separate the triplet and singlet states as shown in Fig. 1(b). The Hamiltonian of the two spin system, $H_0$, can be expressed with the basis set $\{|\uparrow\uparrow\rangle, |\uparrow\downarrow\rangle, |\downarrow\uparrow\rangle, |\downarrow\downarrow\rangle, S_{20}, S_{02}\}$ as [13],

$$H_0 = \begin{bmatrix} E_z/2 & 0 & 0 & 0 & 0 & 0 \\ 0 & E_{z1}/2 & 0 & 0 & t_c & t_c \\ 0 & 0 & -E_{z1}/2 & 0 & -t_c & -t_c \\ 0 & 0 & 0 & -E_z/2 & 0 & 0 \\ 0 & t_c & -t_c & 0 & U_0 - \epsilon & 0 \\ 0 & t_c & -t_c & 0 & 0 & U_0' + \epsilon \end{bmatrix}, \quad (1)$$

where the Zeeman splittings are $E_z \approx \mu_B(g_L B_L + g_R B_R)$, $E_{z1} \approx \mu_B(g_L B_L - g_R B_R)$, in which $B_L = 0.50\ T$ and $B_R = 0.40\ T$ are the magnetic fields and $g_L = 2.00$ and $g_R = 2.00$ are g-factors at left and right quantum dots (QDs) respectively, $\mu_B$ is the Bohr magneton, $t_c$ is the tunnel coupling, $U_0 = U_0' \approx 10\ meV$ is the Hubbard Coulomb interaction energy by assuming the same value for two identical QDs, $\epsilon$ is the detuning energy controlled by the applied detuning gate voltage $\Delta V_g$, which induces the quantum gate operation as schematically shown in Fig. 1(b) [13]. In practice, the g-factors of two QDs can be different and tuned. An estimation of the exchange interaction by using the configuration interaction (CI) calculations [14][15] indicates an extracted tunnel coupling of

$t_c \sim 1\mu eV$ for two QDs 35nm apart. The tunnel coupling is dependent on the interdot distance, barrier height, and materials of the devices, and it is tunable and falls in a wide range of ~0 to ~0.1 $meV$ in practice [16][17]. The value used here is at the lower side. The device is biased in the regime of $U_0 - \epsilon \gg t_c$, in which the Hamiltonian can be simplified to (1).

Two decoherence mechanisms, the charge noise and isotope nuclei dephasing, are modeled. Decoherence due to spin-orbit coupling is weaker and negligible, especially when the external magnetic field direction is optimized [18]. In a MOS device structure, the effect of the interface charge traps is a major limiting factor, and device-to-device variation is an important consideration for large scale integration. To examine the effect of random interface charge traps, the placement of charge traps is determined by a procedure described in Ref. [19], which randomly distributes traps in a region 10,000 times larger than the patterning area of $L_0 \times L_0$ for the two-qubit quantum gate according to a given interface trap density, $N_{IT}$. Here, we assume $L_0 = 100\ nm$ and a state-of-the-art interface quality of $N_{IT} = 2 \times 10^{10} cm^{-2}$ [20], which has an average of 2 traps per individual device.

Charges in the traps induce a random stochastic Coulombic potential, which perturbs the effective detuning potential, as well as the tunnel coupling, and $g$-factor of the spin qubits. While the detuning noise and tunneling noise dominate in different detuning bias conditions, the detuning noise is estimated to be most important for the bias condition examined here [21][22], which is treated in this study. By using the Thomas-Fermi (TF) approximation, the screened trap potential can be expressed as [21],

$$V_{trap}(x,y) = \frac{e^2}{4\pi\varepsilon_{si}[(x-X_T)^2+(y-Y_T)^2]^{3/2}} \left(\frac{1+q_{TF}d}{q_{TF}^2}\right), \quad (2)$$

where $e$ is the elementary electron charge, $\varepsilon_{si}$ is the silicon dielectric constant, $q_{TF} \approx 2/(3nm)$ is the TF screening wave vector, $d \approx 0$ is used for interface traps, $(X_T, Y_T)$ is the in-plane position of a randomly distributed charge trap. To quantify the perturbation, the potential difference between two QDs due to a trap charge is evaluated by computing the expectation value of the screened Coulombic potential, $\langle L|V_{trap}(x,y)|L\rangle - \langle R|V_{trap}(x,y)|R\rangle$, where $|L\rangle$ and $|R\rangle$ are the electron wave functions of the left and right QDs, respectively [23].

Charge traps are assumed to follow the dynamics of random telegraph noise with a characteristic time of $\tau \sim 1ms$, as two level fluctuators. The value of $\tau$ is much larger than the quantum gate time, so the exact value has a small effect on the results. This is a simplified assumption of the charge trap dynamics, and it has been shown that an ensemble of fluctuators with a distribution of time constants can explain experimentally observed nontrivial noise spectrums [24][25][26]. Charge trap dynamics creates a random, time-dependent potential difference between two QDs, which perturbs the detuning energy $\epsilon$ term of the Hamiltonian in Eq. (1).

To model the nuclei dephasing noise, a phenomenological perturbation Hamiltonian $H_{nuc}$ is used. The Overhauser field due to the nuclei spin can be expressed as [27],

$$H_{nuc} = \frac{\mu_B}{2}\sum_{s=L,R} g_s \left(B_{z,s}^{nuc}(t) + \frac{|B_{\perp,s}^{nuc}|^2}{2|B^{ext}|}\right)\hat{\sigma}_s^z = \frac{\mu_B}{2}\sum_{s=L,R}(g_s B_{eff,s})\hat{\sigma}_s^z, \quad (3)$$

where the sum is over the L and R spins, $B_{z,s}^{nuc}(t)$ is the field due to nuclei along the spin direction, $B_{\perp,s}^{nuc}$ is the perpendicular component, $B^{ext}$ is the external field. The combined effect of the field can be described by an effective field $B_{eff,s}$, which can be approximated as a stochastic constant within one quantum gate operation period but varies between different operations. It has been shown that in isotopically purified $^{28}$Si, the spin decoherence time can be >100$\mu s$ [5]. Here the value of $B_{eff,s}$ is phenomenologically determined by requiring the resulting nuclei spin dephasing time to be equal to $100\mu s$.

Based on the system Hamiltonian and description of dephasing mechanisms, a quantum trajectories method (QTM) [28] is used to describe the stochastic gate evolution. The numerical QTM allows incorporate of multiple dephasing mechanisms in a microscopic way, tracks the random trajectories of the propagator, describes the non-Markovian features of the quantum evolution, and removes certain approximations such as perturbative expansion and Gaussian averaging used in analytical models [29][30]. The physical quantities of interest can be derived by statistically averaging over the quantum trajectories. The propagator fidelity can be computed as the expectation value over trajectories, $\mathcal{F} = \langle\frac{|Tr(U_n U_i^+)|}{Tr(U_i U_i^+)}\rangle$, where $U_n$ and $U_i$ are noisy and ideal propagators, respectively [31].

### III. RESULTS AND DISCUSSIONS

The distribution of random interface charge traps is examined first. Fig. 2(a) shows the statistical histogram of the number of interface charge traps per device. While the mean value of traps per device is 2, the device-to-device variation is large. In the 10000 random sampled devices, ~14% have 0 interface traps, while ~6% have ≥5 interface traps. Furthermore, not only does the number of interface charge traps per device vary significantly, the positions of the interface charge traps are randomly distributed, as shown in Fig. 2(b).

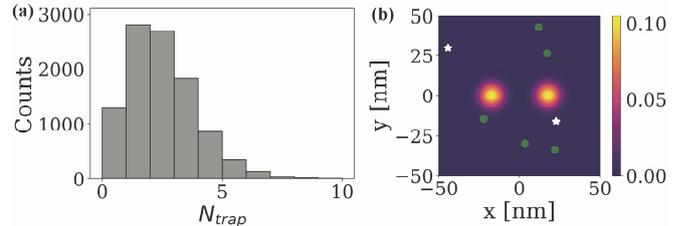

Fig. 2. Random distribution of the interface charge traps. (a) Statistical histogram of the numbers of interface charge traps per device. The area of an individual device is $100nm \times 100nm$, and the interface charge trap density is $N_{IT} = 2 \times 10^{10} cm^{-2}$. (b) shows two device samples of the interface charge trap distributions together with the wave function probabilities of two spin qubits. The device sample 1 have 2 traps and device sample 2 have 5 traps, whose locations are indicated by the white stars and green dots, respectively.

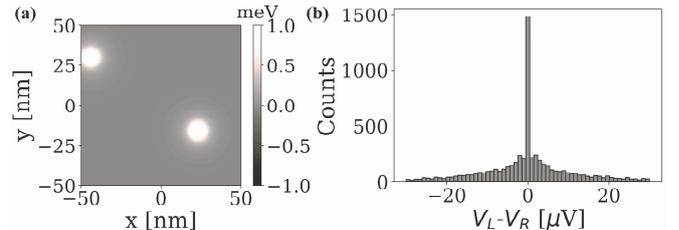

Fig. 3. (a) A sample of the screened charge trap potential profile, which corresponds to the device sample 1 in Fig. 2(b). (b) Statistical distribution of the interdot potential due to charge trapsm which include 10,000 random samples. The modeled device structure is in Fig. 1(a).

Next, the effect of the interface charge traps on the detuning potential is examined. Fig. 3a shows a sample of the screened Coulomb potential profile in the horizontal plane. Because the positions and the numbers of the charge traps distribute randomly, the resulting potential difference, $V_L - V_R$, where $V_L$ and $V_R$ are the expectation values of the charge-trap-induced potentials on the left and right dots, respectively, has a wide distribution, as shown in Fig. 3b. For the statistical device samples that happen to have no interface traps, the potential difference is 0. In addition, those samples with nearly symmetric charge trap distributions with regard to two QDs have small values of $V_L - V_R$. These two cases together result in a peak near 0 in the statistical histogram. For other cases, $V_L - V_R$ ranges from several $\mu V$ to tens of $\mu V$. The results indicate that the perturbation to the detuning potential varies considerably, depending on the numbers and positions of the charge traps.

The effect of the random charge traps on a two-qubit controlled Z (CZ) gate operation realized by the device in Fig. 1(a) is investigated next. The detuning gate operation as shown in Fig. 1(b) differs from the controlled-phase gate locally by single-qubit operations. Because the single-qubit operations have significantly higher fidelity and shorter gate time in silicon qubit technologies, we assume that the single-qubit operations are ideal and ignore their impact on fidelity, which is valid even when the device is in the (1,1) charge stability region if carefully designed pulses are used [10][32]. Fig. 4(a) shows the time evolution of a quantum state for different device samples with random interface traps, in which the left qubit is initialized at $|1\rangle$ and right qubit is initialized at $|+\rangle$. The probability of the right qubit at $|+\rangle$ is shown in Fig. 4(a) as a function of time, which shows device-to-device variations. The device is biased at detuning $U_0 - \epsilon = 0.14\ meV$, which results in a CZ gate time of ~150 ns. Biasing the device closer to the resonant point with $t_c \ll 0.14\ meV$ shortens the gate time, but it does not improve the fidelity, because the effect of detuning noise also increases [22]. Fig. 4(b) plots the statistical histogram of the gate fidelity, which shows a fidelity value of >98% for 75% of statistical device samples, with an average value of 98.8%. The variability of the fidelity between devices, however, is large.

It has been suggested that the fidelity of the two-qubit CZ gates can be improved by using a composite gate pulse [33][34], which suppresses exchange coupling noise regardless of its tunneling- or detuning-caused origin. Therefore, the variability and fidelity limits of the composite quantum gate are examined. The composite gate uses a sequence of two-qubit and single-qubit operations, as shown by the quantum circuit in Fig. 5(a), which is protected against the quasi-static and low-frequency noise for a specific two-qubit noise channel [33][34]. The gates are defined as $S(\Theta) = exp(-i\Theta\sigma_z \otimes \sigma_z/4)$, $R(\theta) = exp(-i\theta\sigma_x/2)$, and the rotational angles are $\theta = \pi - \theta^*$ where $\theta^* \approx 0.674$, $\Theta = -\pi sec(\theta) \approx 1.28\pi$, and $\Theta_2 = 2\pi$ [33][34]. Fig. 5(b), which plots the histogram of the fidelity of the composite gate, shows that the fidelity can be improved to >99.5% for 75% of samples, with an average value of 99.4%. It is noted that the composite gate has more gate stages and about 4.6 times longer two-qubit gate operation time (~700 ns), determined by adding up three two-qubit gate operations with the rotation angles required by the robust design [33][34]. The longer time results in larger impact by nuclei dephasing, which is not protected by the pulse design. Isotopically purified silicon, therefore, is essential for high fidelity of the composite gate. In addition, the composite gate pulse is designed to be ideal for the small noise of the two-qubit ZZ operation channel, based on a perturbative theory approach. The perturbation to detuning potential, as indicated by Fig. 3, however, has a wide distribution. For those device samples with large perturbations due to charge traps, the fidelity value does not show improvement. As a result, although the average of the gate fidelity improves, the variation is still large in the composite quantum gate scheme.

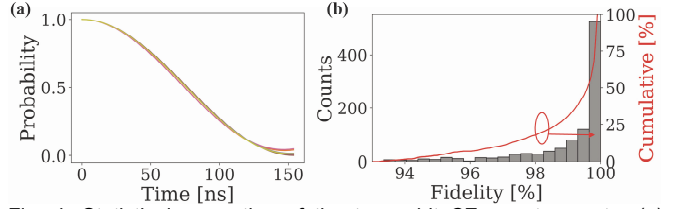

Fig. 4. Statistical properties of the two-qubit CZ quantum gate. (a) Transient characteristics of the CZ gates. The probability of the $|+\rangle$ state of the right qubit is shown, with left qubit at the state '1'. Ten statistical samples are plotted. (b) Histogram bars (left axis) and cumulative curve (right axis) of the fidelity of 1000 random quantum gate samples. The modeled device structure is as shown in Fig. 1(a) with the same interface trap density as Fig. 2.

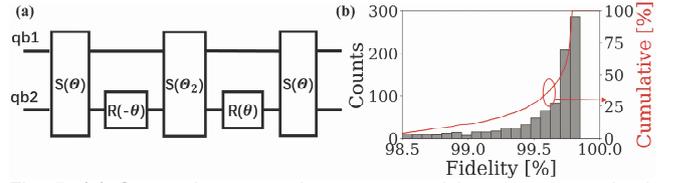

Fig. 5. (a) Composite gate pulse represented by a quantum circuit diagram. The two-qubit ZZ gates are $S(\Theta)$ and $S(\Theta_2)$ with rotation angles of $\Theta$ and $\Theta_2$, respectively. The single-qubit gates $R(\theta)$ and $R(-\theta)$ are rotational gates around $x$ for the right qubit with angles of $\theta$ and $-\theta$, respectively. (b) Histogram bars (left axis) and cumulative curve (right axis) of the fidelity for the composite gate as shown in (a). The three two-qubit stages have the same parameters as in Fig. 4, and the single qubit operations are assumed to be ideal.

## IV. CONCLUSIONS

A simulation method that treats variability of silicon quantum gate devices due to random interface charge trap distributions is developed. The results show that the silicon MOS quantum gates have the potential to achieve the fidelity of >98% with 75% probability, with the requirement of the state-of-the-art MOS interface quality, isotopically purified silicon, and scaled device size for strong interdot coupling. The fidelity can be further improved to >99.5% with 75% probability by using a carefully designed composite gate pulse. The variability between devices, however, is largely due to the small device size and stochastic nature of atomistic scale defects.

## REFERENCES

[1] B. E. Kane, "A silicon-based nuclear spin quantum computer," *Nature*, vol. 393, no. 6681, pp. 133–137, May 1998, doi: 10.1038/30156.
[2] M. A. Eriksson, M. Friesen, S. N. Coppersmith, R. Joynt, L. J. Klein, K. Slinker, C. Tahan, P. M. Mooney, J. O. Chu, and S. J. Koester, "Spin-

Based Quantum Dot Quantum Computing in Silicon," *Quantum Information Processing*, vol. 3, no. 1, pp. 133–146, Oct. 2004, doi: 10.1007/s11128-004-2224-z.

[3] R. Maurand, X. Jehl, D. Kotekar-Patil, A. Corna, H. Bohuslavskyi, R. Laviéville, L. Hutin, S. Barraud, M. Vinet, M. Sanquer, and S. De Franceschi, "A CMOS silicon spin qubit," *Nature Communications*, vol. 7, p. 13575, Nov. 2016, doi: 10.1038/ncomms13575.

[4] M. Veldhorst, H. G. J. Eenink, C. H. Yang, and A. S. Dzurak, "Silicon CMOS architecture for a spin-based quantum computer," *Nature Communications*, vol. 8, no. 1, p. 1766, Dec. 2017, doi: 10.1038/s41467-017-01905-6.

[5] M. Veldhorst, C. H. Yang, J. C. C. Hwang, W. Huang, J. P. Dehollain, J. T. Muhonen, S. Simmons, A. Laucht, F. E. Hudson, K. M. Itoh, A. Morello, and A. S. Dzurak, "A two-qubit logic gate in silicon," *Nature*, vol. 526, no. 7573, pp. 410–414, Oct. 2015, doi: 10.1038/nature15263.

[6] D. M. Zajac, A. J. Sigillito, M. Russ, F. Borjans, J. M. Taylor, G. Burkard, and J. R. Petta, "Resonantly driven CNOT gate for electron spins," *Science*, vol. 359, no. 6374, pp. 439–442, Jan. 2018, doi: 10.1126/science.aao5965.

[7] T. F. Watson, S. G. J. Philips, E. Kawakami, D. R. Ward, P. Scarlino, M. Veldhorst, D. E. Savage, M. G. Lagally, M. Friesen, S. N. Coppersmith, M. A. Eriksson, and L. M. K. Vandersypen, "A programmable two-qubit quantum processor in silicon," *Nature*, vol. 555, no. 7698, pp. 633–637, Mar. 2018, doi: 10.1038/nature25766.

[8] G. Burkard, D. Loss, and D. P. DiVincenzo, "Coupled quantum dots as quantum gates," *Phys. Rev. B*, vol. 59, no. 3, pp. 2070–2078, Jan. 1999, doi: 10.1103/PhysRevB.59.2070.

[9] X. Wang, A. R. Brown, Binjie Cheng, and A. Asenov, "Statistical variability and reliability in nanoscale FinFETs," in *2011 International Electron Devices Meeting*, Dec. 2011, pp. 5.4.1-5.4.4, doi: 10.1109/IEDM.2011.6131494.

[10] W. Huang, C. H. Yang, K. W. Chan, T. Tanttu, B. Hensen, R. C. C. Leon, M. A. Fogarty, J. C. C. Hwang, F. E. Hudson, K. M. Itoh, A. Morello, A. Laucht, and A. S. Dzurak, "Fidelity benchmarks for two-qubit gates in silicon," *Nature*, vol. 569, no. 7757, pp. 532–536, May 2019, doi: 10.1038/s41586-019-1197-0.

[11] R. M. Jock, N. T. Jacobson, P. Harvey-Collard, A. M. Mounce, V. Srinivasa, D. R. Ward, J. Anderson, R. Manginell, J. R. Wendt, M. Rudolph, T. Pluym, J. K. Gamble, A. D. Baczewski, W. M. Witzel, and M. S. Carroll, "A silicon metal-oxide-semiconductor electron spin-orbit qubit," *Nature Communications*, vol. 9, no. 1, p. 1768, May 2018, doi: 10.1038/s41467-018-04200-0.

[12] P. Harvey-Collard, R. M. Jock, N. T. Jacobson, A. D. Baczewski, A. M. Mounce, M. J. Curry, D. R. Ward, J. M. Anderson, R. P. Manginell, J. R. Wendt, M. Rudolph, T. Pluym, M. P. Lilly, M. Pioro-Ladrière, and M. S. Carroll, "All-electrical universal control of a double quantum dot qubit in silicon MOS," in *2017 IEEE International Electron Devices Meeting (IEDM)*, Dec. 2017, pp. 36.5.1-36.5.4, doi: 10.1109/IEDM.2017.8268507.

[13] T. Meunier, V. E. Calado, and L. M. K. Vandersypen, "Efficient controlled-phase gate for single-spin qubits in quantum dots," *Phys. Rev. B*, vol. 83, no. 12, p. 121403, Mar. 2011, doi: 10.1103/PhysRevB.83.121403.

[14] T. Wu, and J. Guo, "Computational Assessment of Silicon Quantum Gate Based on Detuning Mechanism for Quantum Computing," *IEEE Transactions on Electron Devices*, vol. 65, no. 12, pp. 5530–5536, Dec. 2018, doi: 10.1109/TED.2018.2876355.

[15] T. Wu, and J. Guo, "Performance Assessment of Resonantly Driven Silicon Two-Qubit Quantum Gate," *IEEE Electron Device Letters*, vol. 39, no. 7, pp. 1096–1099, Jul. 2018, doi: 10.1109/LED.2018.2835385.

[16] F. A. Zwanenburg, A. S. Dzurak, A. Morello, M. Y. Simmons, L. C. L. Hollenberg, G. Klimeck, S. Rogge, S. N. Coppersmith, and M. A. Eriksson, "Silicon quantum electronics," *Rev. Mod. Phys.*, vol. 85, no. 3, pp. 961–1019, Jul. 2013, doi: 10.1103/RevModPhys.85.961.

[17] U. Mukhopadhyay, J. P. Dehollain, C. Reichl, W. Wegscheider, and L. M. K. Vandersypen, "A 2 × 2 quantum dot array with controllable inter-dot tunnel couplings," *Appl. Phys. Lett.*, vol. 112, no. 18, p. 183505, Apr. 2018, doi: 10.1063/1.5025928.

[18] R. Ferdous, K. W. Chan, M. Veldhorst, J. C. C. Hwang, C. H. Yang, H. Sahasrabudhe, G. Klimeck, A. Morello, A. S. Dzurak, and R. Rahman, "Interface-induced spin-orbit interaction in silicon quantum dots and prospects for scalability," *Phys. Rev. B*, vol. 97, no. 24, p. 241401, Jun. 2018, doi: 10.1103/PhysRevB.97.241401.

[19] Hon-Sum Wong, and Yuan Taur, "Three-dimensional 'atomistic' simulation of discrete random dopant distribution effects in sub-0.1 um MOSFET's," in *Proceedings of IEEE International Electron Devices Meeting*, Dec. 1993, pp. 705–708, doi: 10.1109/IEDM.1993.347215.

[20] D. Bauza, "Extraction of Si-SiO2 interface trap densities in MOS structures with ultrathin oxides," *IEEE Electron Device Letters*, vol. 23, no. 11, pp. 658–660, Nov. 2002, doi: 10.1109/LED.2002.805008.

[21] D. Culcer, X. Hu, and S. Das Sarma, "Dephasing of Si spin qubits due to charge noise," *Applied Physics Letters*, vol. 95, pp. 073102–073102, Sep. 2009, doi: 10.1063/1.3194778.

[22] P. Huang, N. M. Zimmerman, and G. W. Bryant, "Spin decoherence in a two-qubit CPHASE gate: the critical role of tunneling noise," *npj Quantum Inf*, vol. 4, no. 1, p. 62, Dec. 2018, doi: 10.1038/s41534-018-0112-0.

[23] Q. Li, Ł. Cywiński, D. Culcer, X. Hu, and S. Das Sarma, "Exchange coupling in silicon quantum dots: Theoretical considerations for quantum computation," *Phys. Rev. B*, vol. 81, no. 8, p. 085313, Feb. 2010, doi: 10.1103/PhysRevB.81.085313.

[24] J. Yoneda, K. Takeda, T. Otsuka, T. Nakajima, M. R. Delbecq, G. Allison, T. Honda, T. Kodera, S. Oda, Y. Hoshi, N. Usami, K. M. Itoh, and S. Tarucha, "A quantum-dot spin qubit with coherence limited by charge noise and fidelity higher than 99.9%," *Nature Nanotechnology*, vol. 13, no. 2, p. 102, Feb. 2018, doi: 10.1038/s41565-017-0014-x.

[25] U. Güngördü, and J. P. Kestner, "Indications of a soft cutoff frequency in the charge noise of a Si/SiGe quantum dot spin qubit," *Phys. Rev. B*, vol. 99, no. 8, p. 081301, Feb. 2019, doi: 10.1103/PhysRevB.99.081301.

[26] E. J. Connors, J. Nelson, H. Qiao, L. F. Edge, and J. M. Nichol, "Low-frequency charge noise in Si/SiGe quantum dots," *Phys. Rev. B*, vol. 100, no. 16, p. 165305, Oct. 2019, doi: 10.1103/PhysRevB.100.165305.

[27] I. Neder, M. S. Rudner, H. Bluhm, S. Foletti, B. I. Halperin, and A. Yacoby, "Semiclassical model for the dephasing of a two-electron spin qubit coupled to a coherently evolving nuclear spin bath," *Phys. Rev. B*, vol. 84, no. 3, p. 035441, Jul. 2011, doi: 10.1103/PhysRevB.84.035441.

[28] W. T. Strunz, L. Diósi, and N. Gisin, "Open System Dynamics with Non-Markovian Quantum Trajectories," *Phys. Rev. Lett.*, vol. 82, no. 9, pp. 1801–1805, Mar. 1999, doi: 10.1103/PhysRevLett.82.1801.

[29] J. A. Krzywda, and Ł. Cywiński, "Adiabatic electron charge transfer between two quantum dots in presence of 1/f noise," *Phys. Rev. B*, vol. 101, no. 3, p. 035303, Jan. 2020, doi: 10.1103/PhysRevB.101.035303.

[30] E. Paladino, Y. M. Galperin, G. Falci, and B. L. Altshuler, "1/f noise: Implications for solid-state quantum information," *Rev. Mod. Phys.*, vol. 86, no. 2, pp. 361–418, Apr. 2014, doi: 10.1103/RevModPhys.86.361.

[31] J. A. Jones, "Robust Ising gates for practical quantum computation," *Phys. Rev. A*, vol. 67, no. 1, p. 012317, Jan. 2003, doi: 10.1103/PhysRevA.67.012317.

[32] U. Güngördü, and J. P. Kestner, "Robust implementation of quantum gates despite always-on exchange coupling in silicon double quantum dots," *Phys. Rev. B*, vol. 101, no. 15, p. 155301, Apr. 2020, doi: 10.1103/PhysRevB.101.155301.

[33] T. Ichikawa, U. Güngördü, M. Bando, Y. Kondo, and M. Nakahara, "Minimal and robust composite two-qubit gates with Ising-type interaction," *Phys. Rev. A*, vol. 87, no. 2, p. 022323, Feb. 2013, doi: 10.1103/PhysRevA.87.022323.

[34] U. Güngördü, and J. P. Kestner, "Pulse sequence designed for robust C-phase gates in SiMOS and Si/SiGe double quantum dots," *Phys. Rev. B*, vol. 98, no. 16, p. 165301, Oct. 2018, doi: 10.1103/PhysRevB.98.165301.